# Topological Hall-like anomalies from twisted magnetic exchange spring in $Nd_{0.6}Sr_{0.4}MnO_3$ / $SrRuO_3$ heterostructures

**Mrinaleni R S [1,2], *E P Amaladass [1,2], A. T. Sathyanarayana [1,2], Jegadeesan P [1], R Pandian[1,2], S Amirthapandian [1,2], Pooja Gupta [2,3], S K Rai [2,3], And Awadhesh Mani[1,2]**

[1]Material Science Group, Indira Gandhi Centre for Atomic Research, Kalpakkam, 603102
[2]Homi Bhabha National Institute, Training School Complex, Anushakti Nagar, Mumbai, Maharashtra 400094, India
[3]Synchrotrons Utilisation Section, Raja Ramanna Centre for Advanced Technology, PO RRCAT, Indore, Madhya Pradesh 452013, India

**Corresponding author:* edward@igcar.gov.in

## Abstract

In recent times, there has been a surge of interest in utilizing manganite systems for oxide-based heterostructures that host non-trivial spin textures, primarily driven by the Dzyaloshinskii-Moriya interaction (DMI). In these systems, the Topological Hall Effect (THE) manifests as humps or peaks in anomalous Hall Effect (AHE) loops near the coercive field. Numerous studies have reported THE in ruthenate/iridate and ruthenate/manganite heterostructures. Among the manganites, the $Nd_{0.6}Sr_{0.4}MnO_3$ (NSMO) system remains relatively less explored. The strontium ruthenate ($SrRuO_3$ - SRO) system exhibits excellent structural compatibility with NSMO, making NSMO/SRO heterostructures a model system for investigating non-trivial spin textures at oxide-based interfaces. In this work, we have studied the proximity effect of NSMO with SRO through magnetic and magnetotransport studies, varying the SRO layer thicknesses in the NSMO/SRO heterostructures.

## 1. Introduction

Oxide heterostructures exhibit a plethora of exciting phenomena, such as the formation of 2D electron gas at $SrTiO_3$/$LaAlO_3$ interfaces, tuneable ferroelectricity, exchange bias effects, and spin-orbit interactions, making them intriguing for both fundamental research and practical applications [1]. Among the oxides, the $SrRuO_3$ (SRO) system, in particular, has garnered significant attention recently due to its anomalous Hall Effect (AHE) [2,3]. The AHE changes sign near its Curie temperature (Tc) because of intrinsic contributions from the band structure and Berry curvature effects in SRO [2,4]. Furthermore, studies claim to have observed the topological Hall Effect (THE) in thin films of SRO [3]. The Topological Hall Effect (THE) arises in addition to the ordinary Hall Effect and the AHE, in systems hosting non-collinear spin textures, such as helices, domain walls, and skyrmions [5,6]. During transport through these non-collinear spin structures, the electrons gain a quantum mechanical Berry phase by adiabatically following the spin polarization of the magnetic entities. This Berry phase acts as an emergent magnetic field, altering the electrons' trajectories leading to additional contributions to the AHE. However, anomalies like an artefact THE or THE-like features in AHE can also arise due to the two-component anomalous Hall effect [7]. In SRO thin films, structural variations arising from thickness inhomogeneity, crystallographic defects due to Ru vacancies, stoichiometry variations, and rotations and tilts of the $RuO_2$ octahedron are said to give rise to THE-like contributions in AHE [8]. THE-like features have been observed in SRO thin films with thicknesses less than 10 nm [8,9]. To get more insights into the origin of THE-like features in SRO thin films, the first part of this work investigates the magnetic and transport properties of single-layer SRO films on STO substrates of different thicknesses.

The interfacial Dzyaloshinskii-Moriya interaction (DMI) is believed to contribute to the observation of THE-like features in AHE [10,11]. Thin films and heterostructures naturally host Dzyaloshinskii-Moriya interaction (DMI) due to symmetry breaking at interfaces and surfaces, with materials exhibiting high spin-orbit coupling further enhancing the DMI [10,11]. The presence of DMI leads to canting magnetic moments in the system, giving rise to non-collinear spin textures that result in THE [12]. This has drawn significant attention to heterostructures of exotic SRO with other oxide systems such as strontium iridate ($SrIrO_3$) and manganites [3,13,14]. Numerous investigations claim to have observed THE in ruthenate/iridate and ruthenate/manganite heterostructures [3,13,15]. While the LSMO-SRO interface has been extensively studied for THE and exchange coupling [16–18], the $Nd_{0.6}Sr_{0.4}MnO_3$ (NSMO) system remains comparatively underexplored. In this work, we examine the proximity effect between NSMO and SRO in heterostructures through magnetic and magnetotransport investigations. In addition, we analyze single layers of SRO and NSMO on STO and compare their magnetic and transport properties with SRO/NSMO/STO heterostructures of varying thicknesses to gain insights into the emergence of THE-like features in these oxide heterostructures.

## 2. Experimental methods

Thin films and heterostructures of NSMO and SRO were synthesized using the pulsed laser deposition technique (PLD) with a KrF excimer laser ($\lambda$=248 nm) operating at a fluence of 1 $J/cm^2$ with a repetition rate of 3 Hz. A commercial NSMO pellet was used as the target for the heterostructures, and the SRO target was prepared using the solid-state reaction technique. Single crystals of $SrTiO_3$ with (100) orientation were used as substrates and the substrates were treated for $TiO_2$ termination using the protocol discussed in our previous works [19,20]. The oxygen partial pressure was fixed at 0.36 mbar, and the substrate temperature was maintained at 750 $^oC$ during deposition. In-situ annealing was carried out after deposition inside the PLD chamber for 2 h with an oxygen background pressure of 1 bar. The heterostructures are synthesized with the NSMO layer thickness fixed at 40 nm. The SRO layer of different thicknesses was deposited atop the NSMO layer. Thus, the heterostructure comprises a 40 nm NSMO bottom layer and a top SRO layer ranging from 5 to 80 nm in thickness. Thickness measurements were conducted using X-ray reflectivity (XRR) for single layers. The thickness of the bilayers was verified using cross-sectional scanning electron microscopy (SEM) and correlated with the XRR fittings. It was found that SRO thin films have thicknesses of ~6 nm, ~23 nm, and ~50 nm for 250, 500, and 750 laser shots, respectively. The fitted XRR profiles of single layers and a representative bilayer are illustrated in the Figure 1.

For the structural characterization, XRD measurements were carried out at Engineering Applications Beamline, BL-02, Indus-2 synchrotron source, India. The XRD studies utilized the monochromatic high-resolution mode of the Beamline with a beam energy of 15 keV ($\lambda$ = 0.826 Å). Peaks were indexed based on ICDD data (ICDD numbers: 01-085-6743 and 00-043-0472) [21,22]. The XRR measurements were performed in a BRUKER D8, Lab source XRD setup. The surface morphology of the thin films was determined using scanning electron microscopy (SEM) with crossbeam 340 (from Carl Zeiss). The magnetotransport measurements were performed in a cryo-free magnetoresistance setup by Cryogenics, UK. A SQUID-based vibrating sample magnetometer system from Quantum Design (EverCool) was employed for magnetization measurements.

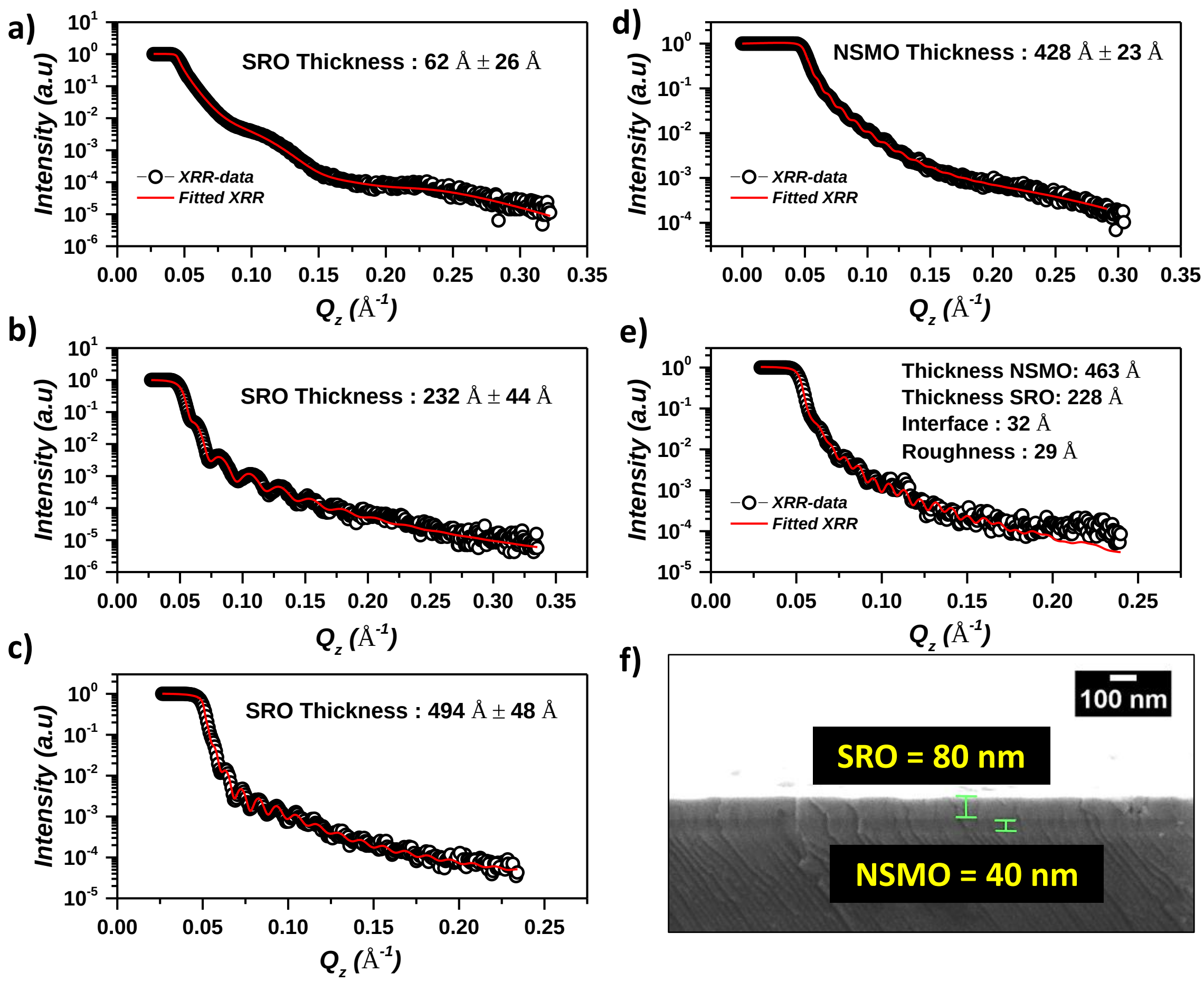

*Figure 1: Thickness determination from XRR fittings using Parratt32 software for SRO, NSMO thin films, and NSMO/SRO heterostructures. XRR fittings of a)-c) SRO thin films, SR-6, SR-23, SR-50 respectively, d) reference NSMO thin film – 40 nm e) NSMO/SRO heterostructure – NS-SR-23 and -f) Thickness of NS-SR-80 determined from cross-sectional SEM, respectively.*

## 3. Results and discussion:

### Magnetic and Magnetotransport properties of Single layer: $SrRuO_3/SrTiO_3$ (100)

SRO thin films of 6 nm, 23 nm, and 50 nm are considered here and referred to as SR-6, SR-23, and SR-50, where the SR represents the SRO layer and the number represents the thickness in nm. To examine the crystallinity and the phase purity of the thin films, XRD scans were performed over the 10° – 50° range in θ-2θ mode, shown in Figure 2 a), and it was found that the films exhibited peaks only corresponding to the out-of-plane reflections. The scans also revealed an absence of impurity peaks, and the films were observed to be highly textured, mirroring the substrate. A high-resolution XRD scan was performed around the (002) STO substrate peak for the SRO thin films, as shown in Figure 2 b)-d). The SRO films crystallize with orthorhombic crystal structure with space group Pnma, and the out-of-plane reflections from the films were indexed as (004) peaks with reference to ICDD data for 15 KeV [22]. The SRO thin films grow in a pseudo-cubic fashion [4,23] strained to the lattice constant of STO and the films experience a compressive strain on STO, which is evident from Figure 2 b)-d) where the SRO peak corresponding to the (004) plane is shifted to the left of the substrate peak. The value of the out-of-plane lattice parameter is found to vary between 8.06 Å for SR-6 and 7.95 Å for SR-50. The decrease in the value of the out-of-plane lattice constant with increase in thickness is due to significant relaxation in compressive strain in films as thickness increases. It is further observed that all three SRO thin films exhibit Laue oscillations, indicating the long-range crystalline order within the thickness of the film with a sharp interface [24].

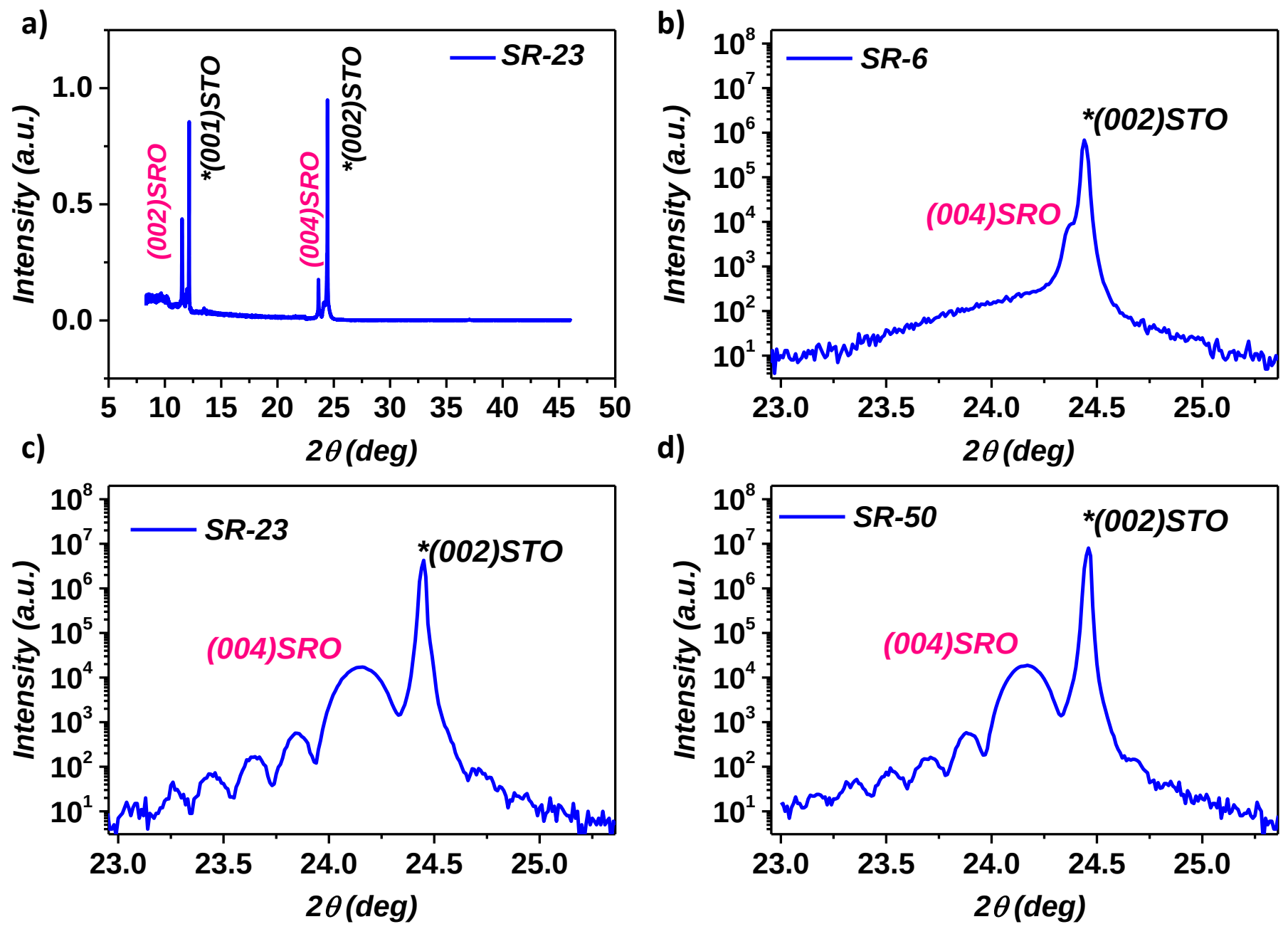

*Figure 2: a) XRD scan of SRO thin film: SR-23 over the full range 5 – 45 deg. HR-XRD scans of SRO thin films: SR-6, SR-23, SR-50 deposited on TiO2 terminated (100)-SrTiO3 substrates with varying thickness exhibiting Laue-oscillations.*

In addition, morphology analysis was carried out using SEM. The morphology of the film was found to possess terrace features originating from the $TiO_2$ terminated STO surface, and the growth is observed along the step edges, indicating a step-flow growth as shown in Figure 3 a). In thicker SRO films grown on STO and NSMO/STO, the terraces were found to vanish, and the films show smooth surfaces with rectangular faceted holes and rod-like features, as shown in Figure 3 b). This is similar to the morphology of a single NSMO layer studied in our previous works [19,20].

The magnetization measurements were carried out on the SR-23 film in two configurations where the H||(001) represents the case with a magnetic field perpendicular to the thin film surface, and H||(100) represents the magnetic field parallel to the thin film surface. Temperature-dependent magnetization, M(T) behavior of SR-23 in the mentioned configurations are shown in Figure 4 a), b). The M(T) shows the magnetic transition of the SRO system from a paramagnetic to a ferromagnetic state. The Curie-temperature is evaluated to be ~ 163 K from the derivative of the field-cooling curve in H||(001) configuration, as shown in Figure 4 c). However, the derivative of the FC data shows an additional peak close to the major peak corresponding to $T_{Curie}$. The derivative plot for different field directions (H||(100) and H||(001)) shows that the peak magnitude changes with the field. Thus indicating the possibility of an additional magnetic phase with reduced Curie temperature and different anisotropy. For instance, the peak in the H||(001) direction at 163 K corresponds to the phase with out-of-plane anisotropy.

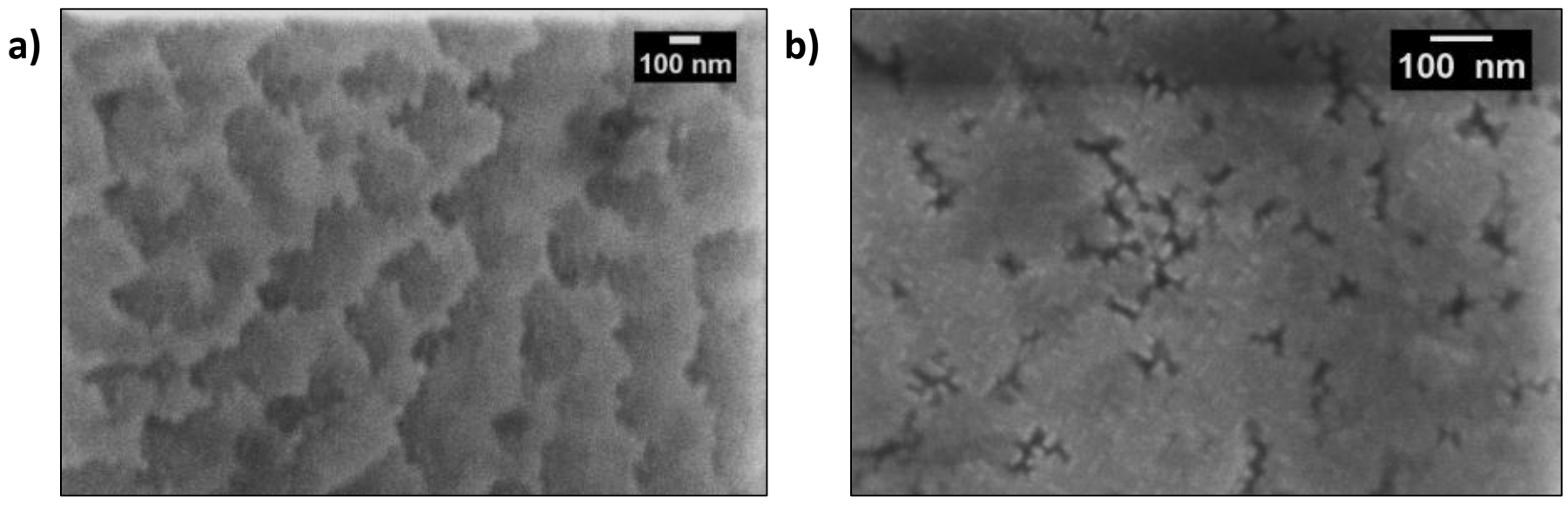

*Figure 3: a) SEM image of 10-nm thick SRO thin film showing terrace-type features b) Surface morphology of NSMO/SRO heterostructure, NS-SR-23, showing rectangular faceted holes and small rod-like features similar to the morphology obtained in previous studies.*

The magnetization hysteresis measured in H||(100) and H||(001) configurations are illustrated in Figure 4 d) and e) where the SRO film shows ferromagnetic hysteresis below 150 K in both configurations and a systematically decreasing coercive field with increasing temperature. A comparison plot of the normalized magnetization hysteresis measured in the in-plane and out-of-plane configurations at 4 K is shown in Figure 4 f). From Figure 4 f), the coercive field measured at 4 K in the H||(100) configuration is around 36 mT, while the coercivity in the H||(001) configuration is 41 mT. The remanence is determined to be 70 emu/cc in the H||(100) configuration, while the H||(001) case shows a remanence of 770 emu/cc. The enhancement in the remanent magnetization, when measured in the H||(001) configuration, indicates that the SRO thin film has an easy axis of magnetization along the out-of-plane direction. It is inferred from the comparison plot that the SRO thin films exhibit out-of-plane (OP) magnetic anisotropy, which matches with the reported claim of literature that SRO thin films grown on (100) STO substrates have an easy-axis tilted about $30^{o}$ degrees away from the surface normal of the film[4,25]. This is because the SRO thin films grow on pseudo-cubic lattice matching with the (100) oriented STO substrate experience compressive strain, and Bruno's model predicts out-of-plane magnetocrystalline anisotropy for this case[26].

Further, it is also noted that the hysteresis measured in H|| (001) configuration is shifted towards the negative quadrant of field. As depicted in Figure 4 g) at 140 K, a complete shift in the hysteresis is observed when measured in H||(001) configuration. However, no such shifts are observed when measured along H||(100). A shift in hysteresis is often attributed to the exchange bias effect, which arises due to uncompensated spins at the AFM/FM interface. The exchange bias effect is usually observed in the magnetization hysteresis after using a field cooling (FC) protocol, where the uncompensated spins are saturated by cooling the sample in high fields prior to measuring the magnetization hysteresis[17]. However, no such FC protocol has been followed in our case.

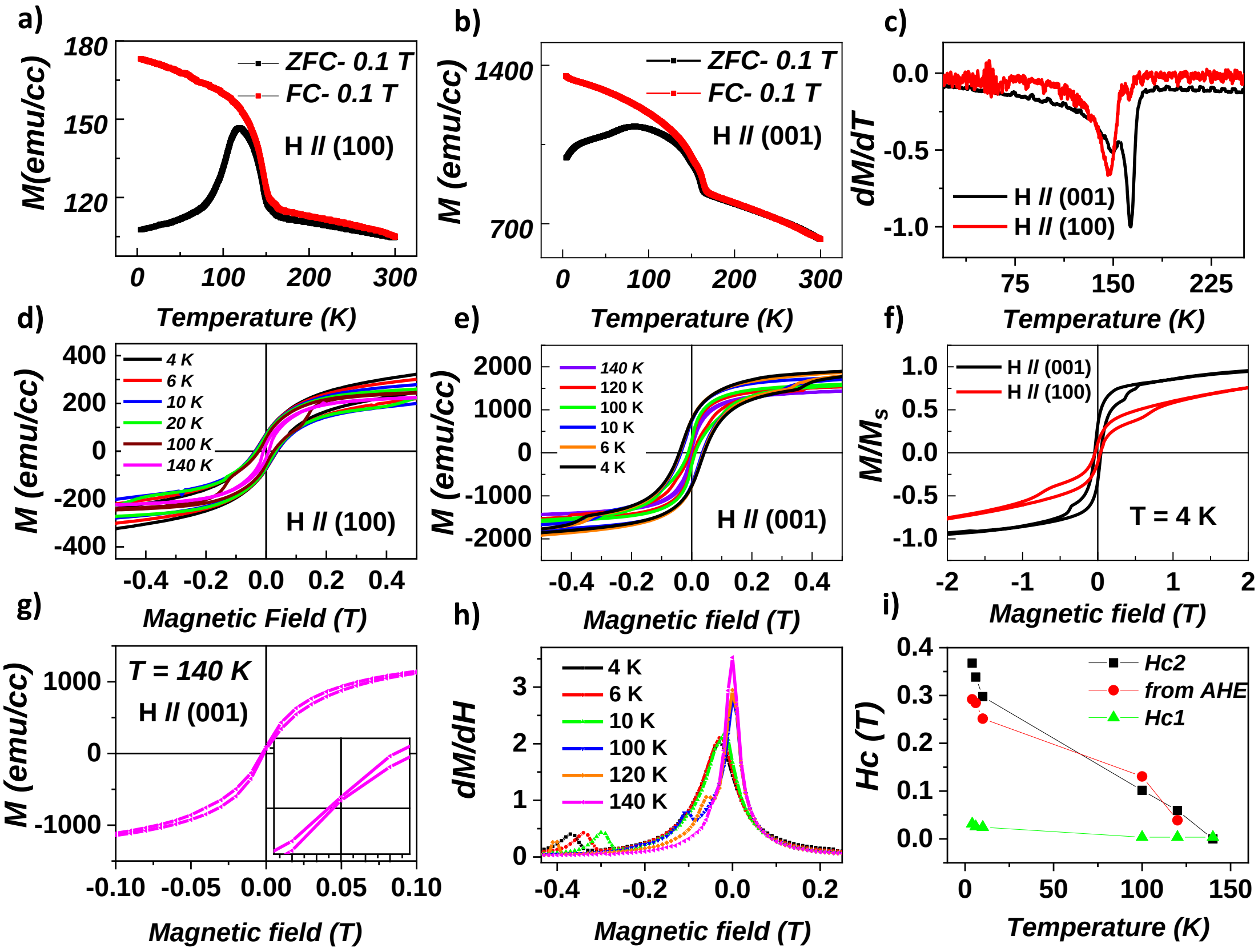

*Figure 4: Temperature-dependent magnetization curve of SR-23 thin film a) measured in the configuration: In-plane - H || (100) and b) out-of-plane - H || (001). c) A derivative plot of the magnetization curve was measured in both configurations, showing two peaks corresponding to different transition temperatures. d), e) Magnetization hysteresis measured in configurations H || (100) and H || (001) f) Normalized magnetization at 4 K the SRO thin film exhibits perpendicular magnetic anisotropy. g) Shift in magnetization hysteresis when measured in H || (001) configuration at 140 K. h) Derivative of magnetization for a single sweep of the magnetic field from + saturation to – ve saturation, exhibiting multiple jumps i) Plot showing the presence of different coercive fields present in the SRO system.*

Additionally, the comparative plot of magnetization at 4 K clearly shows hump-like features at fields 0.35 T in H||(001) configuration and 0.7 T in the case of H||(100) symmetrically in both the sweep directions. Let us refer to the field at which hump occurs

in higher fields as $H_{C2}$. A closer inspection reveals that the humps occur during the reversal of field from saturation and manifest as a jump in magnetization upon reaching the $H_{C2}$, -0.35 T in H||(001) and -0.7 T in H||(100). Again, after saturation at – 3T, the field is reversed towards + 3T, and the hump is observed as an increase in magnetization at +0.35 T in H||(001) and +0.75 T in H||(100) configuration. The humps in magnetization indicate the presence of two magnetic phases with different coercivities. To confirm this, the derivative of magnetization is taken for the sweep from -3 T to 3 T and plotted in Figure 4 h). The differential curve dM/dH shows two peaks, with $H_{c1}$ corresponding to the coercivity Hc and the other close to the $H_{c2}$. This evidently verifies that SRO thin film has two magnetic phases, one with a high Hc and the other with a one-order lower in magnitude. The behavior of Hc as a function of temperature is plotted in Figure 4 i). A decreasing trend in the $H_C$ is evident with increasing in temperature. Thus, the SRO thin films possess two-magnetic phases and exhibit a two-step magnetization reversal. We further investigate the resistivity and magnetotransport behavior of the SRO thin films.

Transport measurements of the 6, 23, and 50 nm SRO thin films are shown in Figure 5 a). R(T) data shows a metallic trend with discernable slope change at $T_{Curie}$. The residual-resistivity ratio (RRR= R (300 K)/R (5 K)) of the film was found to be RRR ≈ 4. Though relatively low, from literature, an RRR in the range of 3.5 to 5 indicates the good crystalline quality of SRO thin films. The presence of a large density of point defects and vacancies could be the reason for the lower RRR values [15]. These defects include Ru deficiency, oxygen vacancies, and modifications to the $RuO_2$ octahedron's rotations and tilts. The magnetotransport measurements were performed over the temperature range from 4.2 K to 150 K on the thin films of SRO. Figure 5 b)-d), f)-h) shows the plot of anomalous resistivity and longitudinal magnetoresistance (MR %) of the SRO thin films. The hall data recorded during the magnetotransport measurements can be expressed in terms of different contributions given by $\rho_{yx} = \rho_{OHE} + \rho_{AHE} + \rho_{THE}$ which includes ordinary hall component, anomalous hall contribution and topological hall component (if any) [9,16]. The ordinary hall component is given by $\rho_{OHE} = R_o H_z$, where $R_o$ represents the hall-coefficient and Hz is the applied transverse magnetic field, and anomalous hall component is expressed as $\rho_{AHE} = R_A M_z$ since it is proportional to the component of magnetization along the applied magnetic field direction and $R_A$ is the anomalous hall coefficient [9,16]. The anomalous hall component is extracted by anti-symmetrization of the hall data collected from the hall measurements, followed by the subtraction of the ordinary hall component ($\rho_{OHE}$).

All three films show anomalous Hall Effect (AHE) with a negative ordinary-hall background due to the majority carriers of electrons in the SRO system. The AHE loop's sign changes around 120 K to 140 K across the SRO thin films, which is evident from the sign change of $\rho_{AHE}$ plotted vs temperature in Figure 5 e) (value of $\rho_{AHE}$ is taken to be the value at saturation). The AHE generally has contributions from both intrinsic and extrinsic mechanisms. The extrinsic mechanisms involve scattering effects such as side jumps and skew scattering. At the same time, the intrinsic AHE arises from the Berry phase mechanism, which is believed to cause the change in AHE in the SRO system around its Curie temperature [27,28]. Thus, the change in the sign of $\rho_{AHE}$ clearly indicates that SRO thin films have intrinsic contributions to their AHE.

The magnetoresistance plots, as seen in Figure 5 f)-h) shows that the films exhibit a negative magnetoresistance as well as peaks in MR near the coercive fields, which is typical for their FM characteristics. At T = 4.2 K, the MR % evaluated at 3 T for SR-6 is -2.3 %, for SR-23 is -1.6 %, and for SR-50 is -2.1 %. The higher MR % at the lowest thickness is due to increased scattering effects in the thinner SRO film. This is also evident from the large coercivity of AHE in the thinnest SRO film SR-6 due to increased pinning centers. Additionally, it is observed that the $H_c$ derived from AHE closely matches with MR's peaks. The value of $H_c$ derived from AHE is also plotted to the $H_c$ comparison plot in Figure 4 i). However, the value of $H_c$ derived from AHE is close to that of $H_{c2}$ obtained from magnetization. This is because the magnetic phase with higher coercivity dominates the transport properties. The MR and hall transport of SRO single layers studied comprehensively revealed the absence of additional THE-like contributions to the AHE as reported in previous reports on ultrathin SRO films [9].

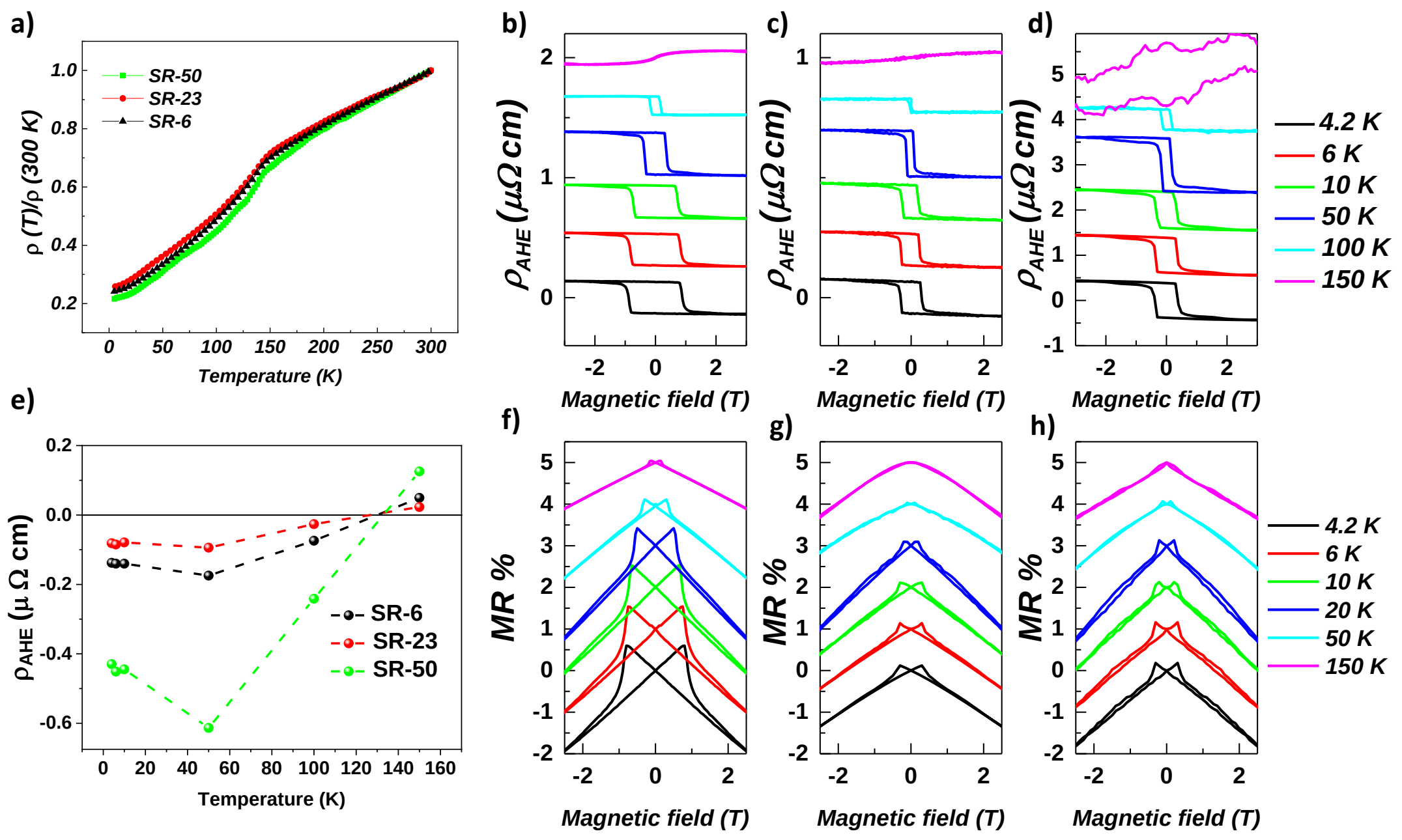

*Figure 5: a) Normalized temperature-dependent resistivity plots of SRO thin films. b)-d) Hall measurements on SRO thin films of thickness 6 nm, 23 nm, and 50 nm, respectively, where the Anomalous Hall Effect is observed across all the SRO thin films without additional dips/humps. e) Evolution of sign change in anomalous hall resistivity with respect to temperature for SRO thin films. f)-h) Corresponding magnetotransport curves of SRO thin films where a negative MR % is obtained across all the films.*

Having ruled out the additional contributions to AHE in SRO single layers, we study the effect of the proximity of NSMO to the SRO system. Apriory to that, the magnetic and transport behavior of the single layer NSMO thin film on STO substrate is investigated in the following section.

**Magnetic and Magnetotransport properties of Single layer: $Nd_{0.6}Sr_{0.4}MnO_3/SrTiO_3$ (100)**

An NSMO thin film of thickness ~ 40 nm is taken as the reference film representing the bottom layer of the heterostructure. From our previous work [19] we have observed that NSMO thin films grown on $TiO_2$-terminated STO substrates grow layer-by-layer, possessing a highly textured structure that mirrors the substrate. A high-resolution XRD scan around the substrate peak (002) shown in Figure 6 a) reveals the presence of (004) out-of-plane reflection of NSMO. The temperature-dependent magnetization in H || (100) configuration in shown in Figure 6 b). The NSMO system shows a magnetic transition from PM to FM at temperatures < 200 K.

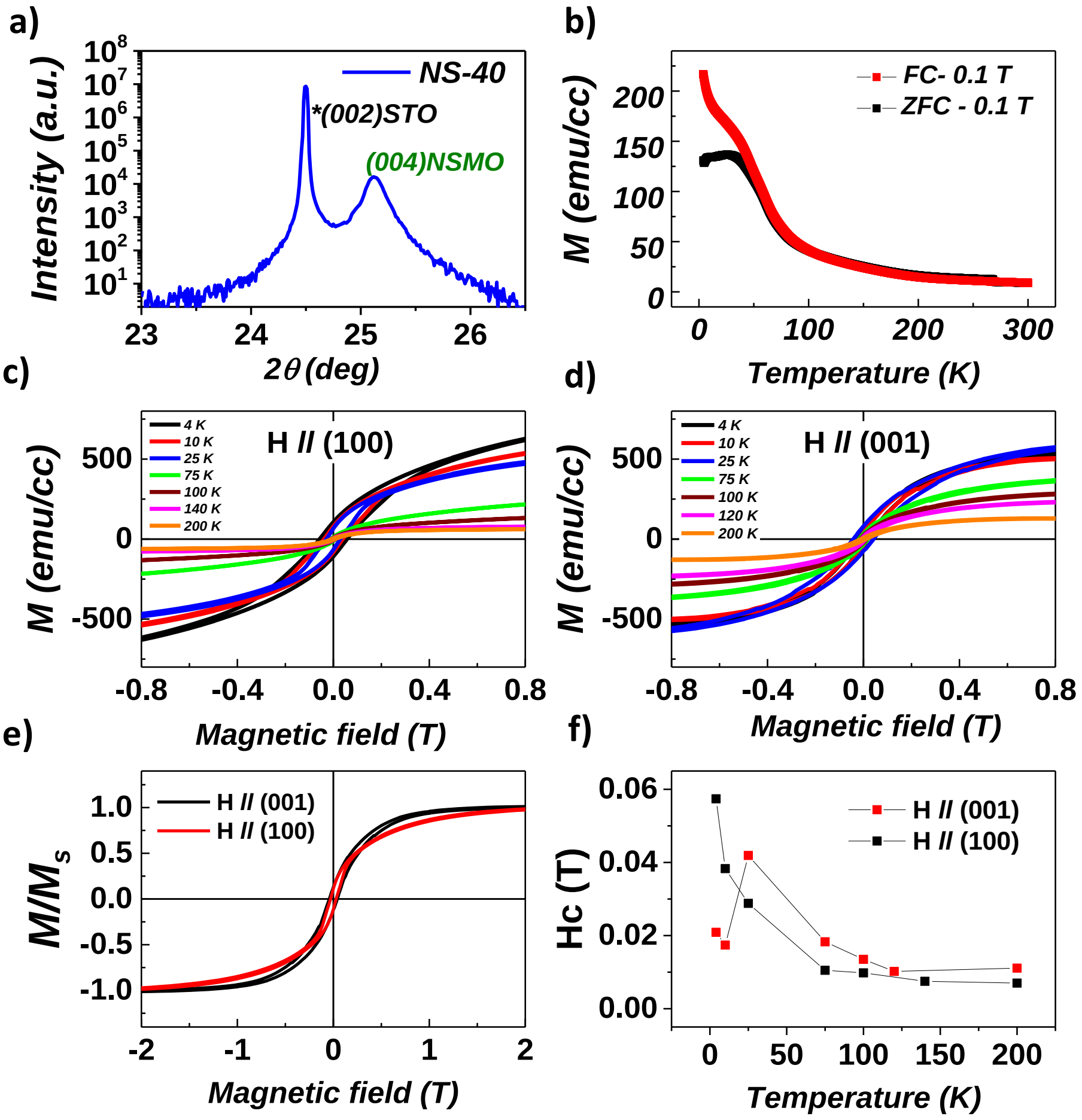

*Figure 6: a) High-resolution XRD scan of NSMO thin film NS-40. b) Temperature-dependent magnetization measured with H || (100) showing FM character below 200 K. c), d) Magnetization hysteresis measured in H || (100) and H || (001) configuration respectively. e) Normalized magnetization hysteresis comparing the magnetization at low-temperature (25 K) when measured with H || (100) and H || (001). f) Temperature dependence of coercive field.*

The in-plane and out-of-plane magnetization hysteresis is illustrated in Figure 6 c), d). The hysteresis measurements reveal a soft magnetic behavior. It is noted that the NSMO thin film, in this case, is not post-annealed. Without post annealing, the NSMO thin films may possess large amount of oxygen deficiency, which predominantly affects the magnetic properties, causing a reduction in the net magnetization as well as deterioration of the magnetic anisotropy of the system. Thus, the NSMO thin film without a strong magnetic anisotropy behaves like a soft magnet. The NSMO thin film exhibits a low coercive field in the range of 10-60 mT and shows a decreasing trend with an increase in temperature, as shown in Figure 6 f). Compared to the coercive field of the SRO thin film, the NSMO thin film's coercivity is lower by an order of magnitude.

The electrical transport measurement revealed that the NSMO thin film's resistance is several MΩ with a negative temperature coefficient of resistance. Below 100 K, its value overshoots the measurement range of the experiment. However, we have seen that ex-situ annealing aids in the reduction of resistance of the sample within the measurable range of the system, thereby exhibiting an insulator-to-metal transition accompanying its magnetic transition[19,20]. Thus, in our case, the NSMO thin film without any post-annealing shows a highly insulating behavior with a soft FM magnetic nature. Therefore, the heterojunction is magnetically coupled, however, the transport measurements are dominated by the SRO layer without redistribution of currents across the individual layers in the heterostructure. This is quite important because many studies have claimed that the redistribution of current across the individual layers like a parallel network leads to two-channel anomalous hall transport giving rise to additional humps and dips in the AHE, which appear like THE[29].

**Magnetic and Magnetotransport properties $SrRuO_3/Nd_{0.6}Sr_{0.4}MnO_3/SrTiO_3$ (100) Heterostructures:**

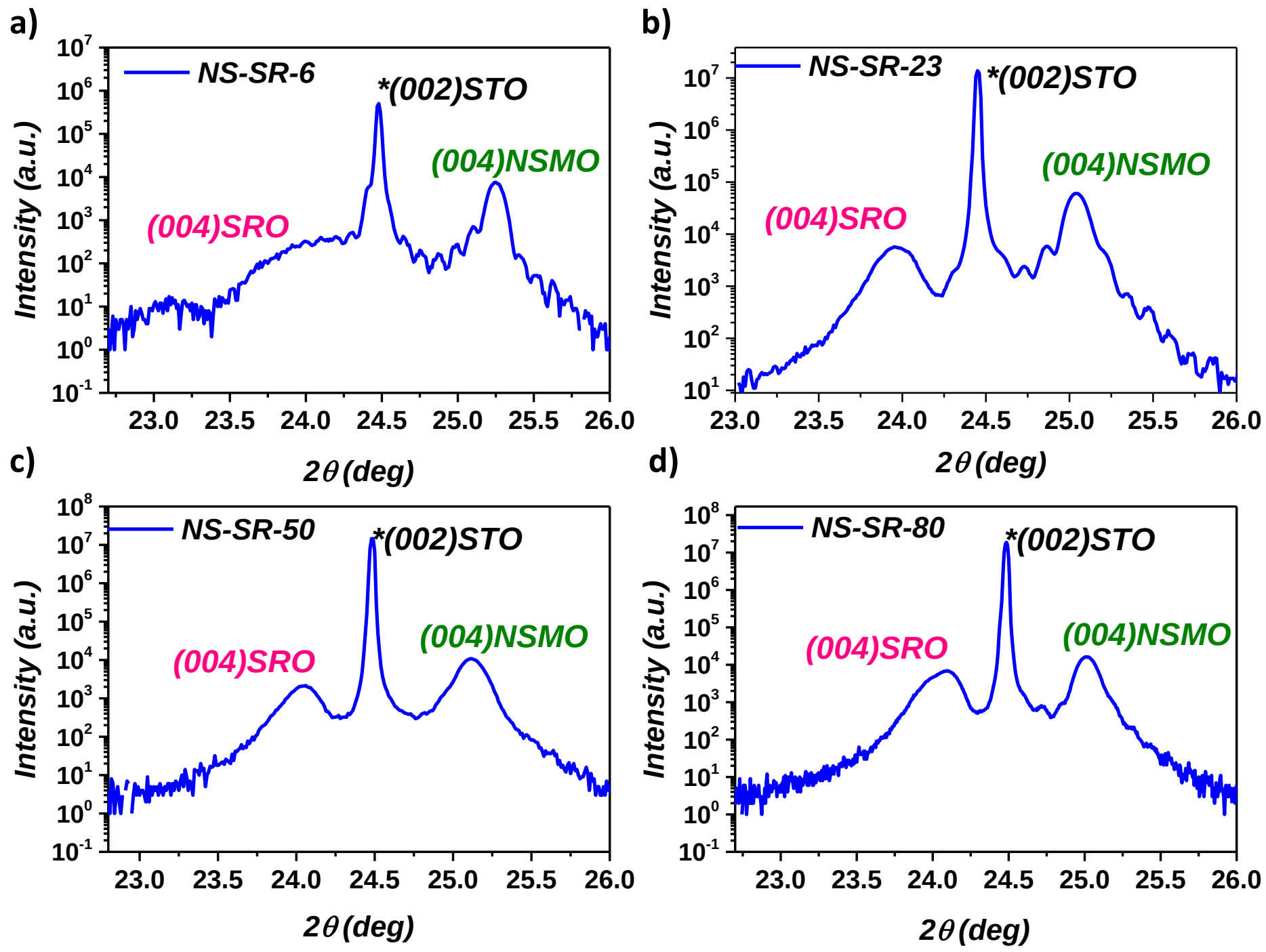

*Figure 7: High-resolution XRD scans of NSMO/SRO heterostructures around (002) substrate peak. a) NS-SR-6 b) NS-SR-23 c) NS-SR-50 d) NS-SR-80.*

The $SrRuO_3/Nd_{0.6}Sr_{0.4}MnO_3/SrTiO_3$ (100) heterostructures were prepared with a constant NSMO thickness of 40 nm and varying SRO layer thickness. We refer to the films as NS-SR-d where d = 6, 23, 50, and 80 nm, representing the thickness of the top SRO layer. Figure 7 presents the high-resolution XRD scans around the substrate peak (002), where the NSMO and SRO layers show reflections along the out-of-plane direction mirroring the substrate. The presence of Laue oscillations, even in the heterostructure with the thinnest SRO layer, indicates that the films are crystalline and homogeneous, with high-quality interfaces between the film and substrate, as well as between the individual layers. A systematic relaxation of compressive strain is observed across the heterostructures with increasing SRO layer thickness, as indicated by the shift in θ position of the SRO peak, as tabulated in Table 1.

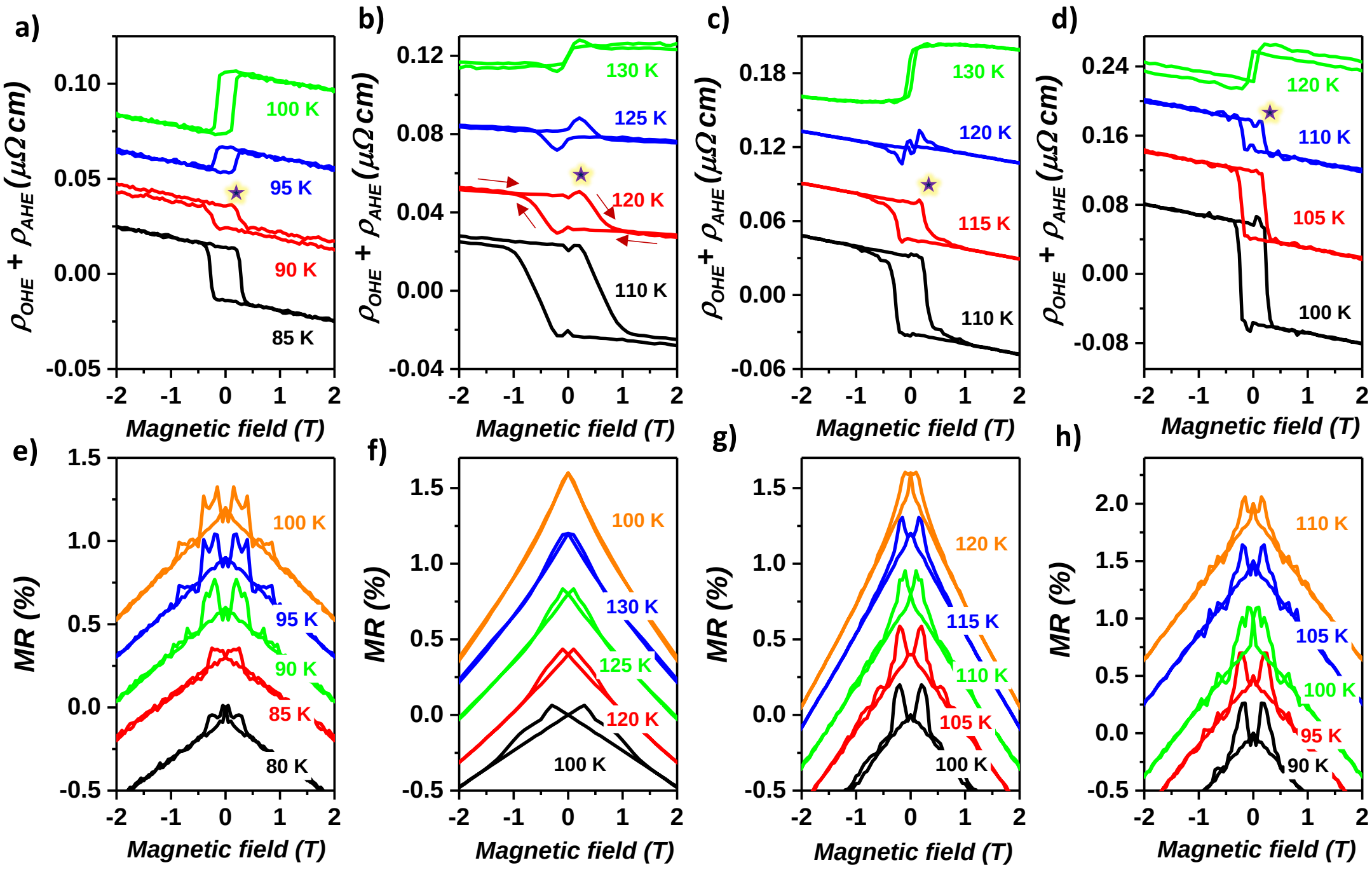

*Figure 8: Magnetotransport measurements on NSMO/SRO heterostructures: NS-SR-6, NS-SR-23, NS-SR-50, and NS-SR-80 a)-d) The corresponding hall resistivity plots where Anomalous Hall Effect (AHE) is observed across all the heterostructures exhibiting sign change in AHE with additional dips/humps near the cross-over temperature similar to Topological Hall effect (Marked with a star). The red arrows at 120 K in b) indicate the occurrence of THE with respect to the direction of field reversal. e)-f) Corresponding linear magnetoresistance plots of NSMO/SRO heterostructures showing negative MR % and additional humps near the MR-peak.*

The magnetotransport properties of the four heterostructures NS-SR-6, NS-SR-23, NS-SR-50, and NS-SR-80 are studied in the upcoming section. Figure 8 illustrates the magnetotransport results of heterostructures. Similar to the single layer SRO, the bilayers exhibit AHE with a negative ordinary hall background, including a temperature-dependent crossover where the sign of the AHE changes, in Figure 8 a)-d). A closer inspection of the AHE data near the temperature at which the sign change occurs reveals an anticipated feature: humps or peaks in the AHE around the coercive field before the crossover. This feature is present across all heterostructures, which is intriguing. The temperature range over which the AHE changes sign varies among the heterostructures. Table 2 lists the temperature range of AHE crossover and the temperature at which the THE-like feature appears. It is observed that the crossover range is lower in the thinner heterostructure (NS-SR-6). However, in the thicker heterostructures, the crossover range is significantly higher but still lower than in the single-layer SRO system. This suggests that interfacial magnetic coupling between the SRO and NSMO layers exists, with its strength varying according to the SRO layer thickness. Moreover, the presence of THE-like hump in the heterostructures and its absence in single-layer SRO clearly indicates that the THE-like feature arises inherently from the proximity of the NSMO layer with SRO.

*Table 1: Illustrates shift in 2θ position based on HR-XRD data, with a systematic increase in thickness of SRO layer.*

| Sample | Position of NSMO (004) peak | Position of SRO (004) peak | Δθ |
|---|---|---|---|
| NS-SR-6 | 25.25 | 23.91 | 1.34 |
| NS-SR-23 | 25.04 | 23.95 | 1.09 |
| NS-SR-50 | 25.11 | 24.06 | 1.05 |
| NS-SR-80 | 25.00 | 24.09 | 0.91 |

The longitudinal magnetoresistance (MR) is also examined, in which, similar to the single-layer SRO, shows a negative magnetoresistive behavior with MR peaks occurring near the coercive fields, as shown in Figure 8 e)-h). Near the temperature where the THE-like feature emerges, the MR plot shows additional humps close to the central MR peak. In comparison, NS-SR-6 shows a small hump near the MR peak, while it is significantly broad in the NS-SR-23 sample. With a further increase in the SRO layer thickness, the hump becomes less discernible. Such additional features in MR at low-fields have been linked to various magnetization processes within the SRO layer, as studied in previous works on LSMO/SRO superlattice [17,18,30,31]. To comprehensively understand these magnetization processes in SRO, we analyze the results of magnetization measurements on the heterostructures. We first study the magnetic properties of the heterostructure NS-SR-23 with intermediate SRO thickness, which gives a prominent THE-like feature.

The temperature-dependent magnetization of NS-SR-23 recorded in two configurations, H||(001) and H||(100), is shown in Figure 9 a),b). In configuration H||(100), the magnetization data shows two transitions at 200 K and 100 K, corresponding to the magnetic transition of the NSMO and SRO systems. When measured with H||(001), the magnetization increases as temperature decreases from 300 K. A jump in magnetization below 150 K indicating the magnetic transition of SRO from PM state to FM state is also evident. Interestingly, at temperatures below 130 K, the net magnetization of the bilayer decreases significantly. Such behavior has been previously observed in LSMO/SRO heterostructures, where the magnetic moments of LSMO and SRO are antiferromagnetically

coupled at the interface [30]. Phenomenologically, the antiferromagnetic (AFM) coupling between $Mn^{3+/4+}$ and $Ru^{4+}$ ions is primarily understood through the super-exchange interaction mediated by oxygen [18,30]. In addition, DFT calculations and XMCD studies confirm a strong AFM exchange coupling between Mn-O-Ru at the interface of LSMO/SRO [18,30,32]. The existence of an interfacial SRO layer ~3.2 nm is evident from our XRR studies, as shown in Figure 1 e). Therefore, once ferromagnetic ordering is established in SRO, the interfacial SRO and NSMO layers couple antiferromagnetically, leading to a net reduction in magnetization below 130 K.

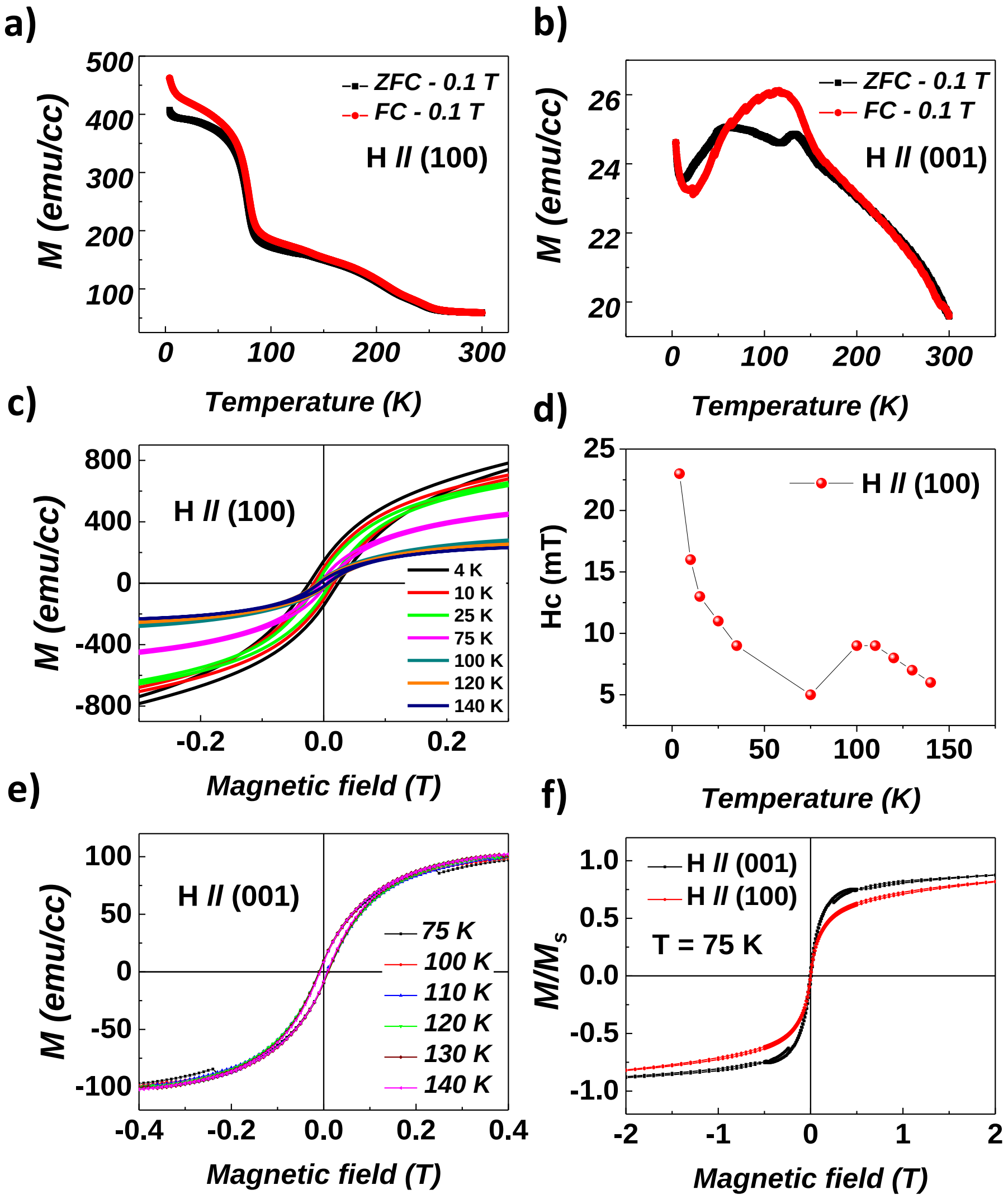

*Figure 9: Temperature-dependent magnetization behavior of NS-SR-23 when measured with magnetic field a) H||(100 ) and b) H||(001). c) Field-dependent hysteresis of the bilayer when H||(100) at 4K – 140 K d) Corresponding temperature dependence of coercivity. e) Magnetization hysteresis measured with H || (001) from 75 K to 140 K showing multiple jumps during reversal. f) Normalized magnetization comparing the hysteresis at 75 K where the bilayer shows magnetic anisotropy along out-of-plane direction.*

Further, comparing the temperature-dependent magnetization values measured in H||(001) and H||(100) configurations, as shown in Figure 6.10, reveals that the magnetization is higher for the H||(100) configurations. Unlike the H||(001) configuration, where a significant reduction in magnetization is observed below 150 K due to the emergent AFM coupling between SRO and NSMO, no substantial decrease in magnetization is observed for H||(100). However, a slope change in magnetization is noted, where the net magnetization changes slowly between 150 K and 130 K. The interplay between the Zeeman energy, exchange coupling, magnetic anisotropy, and their temperature dependence results in such a behavior. Due to the favorable AFM coupling in the H||(001) configuration, the net magnetization drops significantly compared to the H||(100) configuration. A comparative magnetization hysteresis measured at 75 K in both configurations, as shown in Figure 9 f), substantiates the presence of out-of-plane anisotropy in the exchange coupled system.

*Table 2: Tabulation of temperature at which AHE crossover is observed for NSMO/SRO heterostructures*

| Sample | The temperature at which AHE changes sign | The temperature at which THE is observed |
|---|---|---|
| NS-SR-6 | 95 K | 90 K |
| NS-SR-23 | 130 K | 110 K |
| NS-SR-50 | 130 K | 115 K |
| NS-SR-100 | 120 K | 110 K |

The magnetization hysteresis measurements at different temperatures in both configurations are shown in Figure 9 c). The in-plane hysteresis shows a systematic decrease in coercivity from 4 K to 75 K, as depicted in Figure 9 d). However, at 100 K, the coercivity increases before decreasing again with subsequent temperature increases up to 140 K. The magnetization hysteresis measured with the magnetic field applied along the (001) direction (H||(001)) shows prominent jumps. This anomaly can be explained by the exchange-spring coupling between SRO and NSMO. While the AFM coupling across the interface is retained, the reduction in SRO's magnetization with increasing temperature alters the exchange-spring coupling across the SRO/NSMO interface, leading to a non-monotonous change in coercivity as a function of temperature. The prominent jumps observed in the H||(001) measurements can be attributed to the non-coherent switching in the coupled magnetic layers. The presence of such features only when measured along H||(001) also indicates the existence of progressive non-collinear sub-networks of magnetization along the out-of-plane direction. This behavior is analogous to an exchange spring with an out-of-plane component, which has been extensively observed and validated in manganite/ruthenate systems [16,18].

To gain more insight into the coupling across the heterojunction, the hysteresis loops measured with the magnetic field H||(001) are examined in detail. The low-temperature

hysteresis plots, shown in Figure 10, include measurements at T = 4 K, 10 K, 15 K, and 75 K. At 4 K, the hysteresis shows a very low coercivity below 10 mT, indicating the AFM coupling in the system. The hysteresis also appears to be shifted in a negative direction, similar to exchange bias. At higher fields, the individual layers align along the magnetic field direction due to high Zeeman energy, consequently breaking the exchange spring coupling. This is observed as an irreversibility in magnetization, as seen in the hysteresis opening upon reversal of the magnetic field from saturation [17,30]. The presence of small jumps is also noted in the hysteresis. As the temperature is increased to 10 K, the hysteresis becomes more interesting with an unmeasurably low coercivity < 1 mT with a mild negative remanence as the field is reduced from saturation field in Figure 10 b). Following the black arrow, while the field is reversed from a saturation of +1 T, the magnetization keeps decreasing to very low values, and at zero field, the remanence becomes negative (-1.8 emu/cc). With further increasing the field towards the negative direction (-1 T) T, the magnetization exhibits a peculiar jump at 0. 17 T, where the magnetization abruptly drops. With increasing field, saturation is reached at -1 T. Upon reversal of the field from – 1 T to +1 T, a similar signature repeats. The magnetization loop measured at T = 15 K shows an inverted hysteresis for 30 mT > B > 30 mT. As the field is reduced from positive saturation field, the magnetization decreases continuously, and a negative remanence is observed at zero field. With further increase in the field, a jump is noted around 20 – 30 mT, and saturation is reached. Upon reversal, the same characteristics are observed.

In contrast to the low-temperature behavior, at T = 75 K, a magnetization hysteresis is observed with the presence of multiple jumps in magnetization as the field is reduced from saturation. As closer view of the magnetization jumps is shown in Figure 11 b). An increased magnetization emerges with the jumps/steps in the positive quadrant, at +0.48 T and +0.25 T, followed by a plateau region in the negative quadrant at - 0.4 T during reversal. Similar features are also evident upon reversal of the magnetic field from -1 T towards 1 T, Figure 11 c). Complete saturation is observed for fields above 5 T, but magnetization irreversibility is present between 0.5 T to 5 T. The magnetization hysteresis measured at temperatures above 75 K is shown in Figure 11 d). Similar to the 75 K case, the magnetization hysteresis exhibits prominent jumps and magnetization irreversibility. The field at which these jumps occur shows a systematic increase toward higher fields with increase in temperature.

The features observed in the temperature-dependent magnetization reversal curves can be understood through different magnetization reversal mechanisms prevalent in such exchange-spring coupled systems. Previous studies on the magnetization behavior of LSMO-SRO heterostructures and superlattices provide insights for interpreting our results[18,31]. For instance, M. Ziese et al. reported that the LSMO/SRO superlattices exhibit similar magnetization reversal behavior at low temperatures, with an inverted hysteresis and a normal hysteresis at higher temperatures (around 100 K) [31]. As stated by Ziese et al. in the bulk LSMO, the magnetic moment of $Mn^{3+/4+}$ is 3.7 $\mu_B$, while in bulk SRO, the $Ru^{4+}$ moment is 1.6 $\mu_B$, forming a ferrimagnetic network in their superlattice[31]. They explain that the inverted hysteresis at low temperatures arises from the softer LSMO layer switching before the harder SRO layer, whereas at higher temperatures, the SRO layer switches first. Additionally, the detailed switching mechanism is explained from the point of view of Bloch wall formation (exchange spring)[31]. Thus, drawing an analogy from their work, we propose two reversal models to comprehend our magnetization data. Model one represents the reversal for the temperature range of 4 to 75 K, and model two for temperatures above 75 K.

Before establishing the model, some foundational facts regarding the NSMO/SRO system are stated. The NSMO system belongs to the manganite family, thereby possessing a larger magnetic moment as compared to the SRO system, and we expect the formation of a similar ferrimagnetic network across the heterojunction, which aligns with our previous claim of progressive non-collinear sub-networks of magnetization. Additionally, for a given thickness, orthorhombic SRO thin films grown on STO (100) substrates exhibit high magnetocrystalline anisotropy (MCA) with an out-of-plane canting, making them the hard phase at low temperatures. In contrast, manganite thin films on STO have a negative anisotropy constant and in-plane magnetization due to shape anisotropy [18,33,34]. Our previous section shows that single-layer in-situ annealed NSMO is soft magnetic. However, the net magnetization of these layers is influenced by both thickness and temperature. Therefore, the interplay between the magnetization strength and magnetic anisotropy as a function of temperature results in the complex magnetization reversal behavior observed in the NSMO-SRO system.

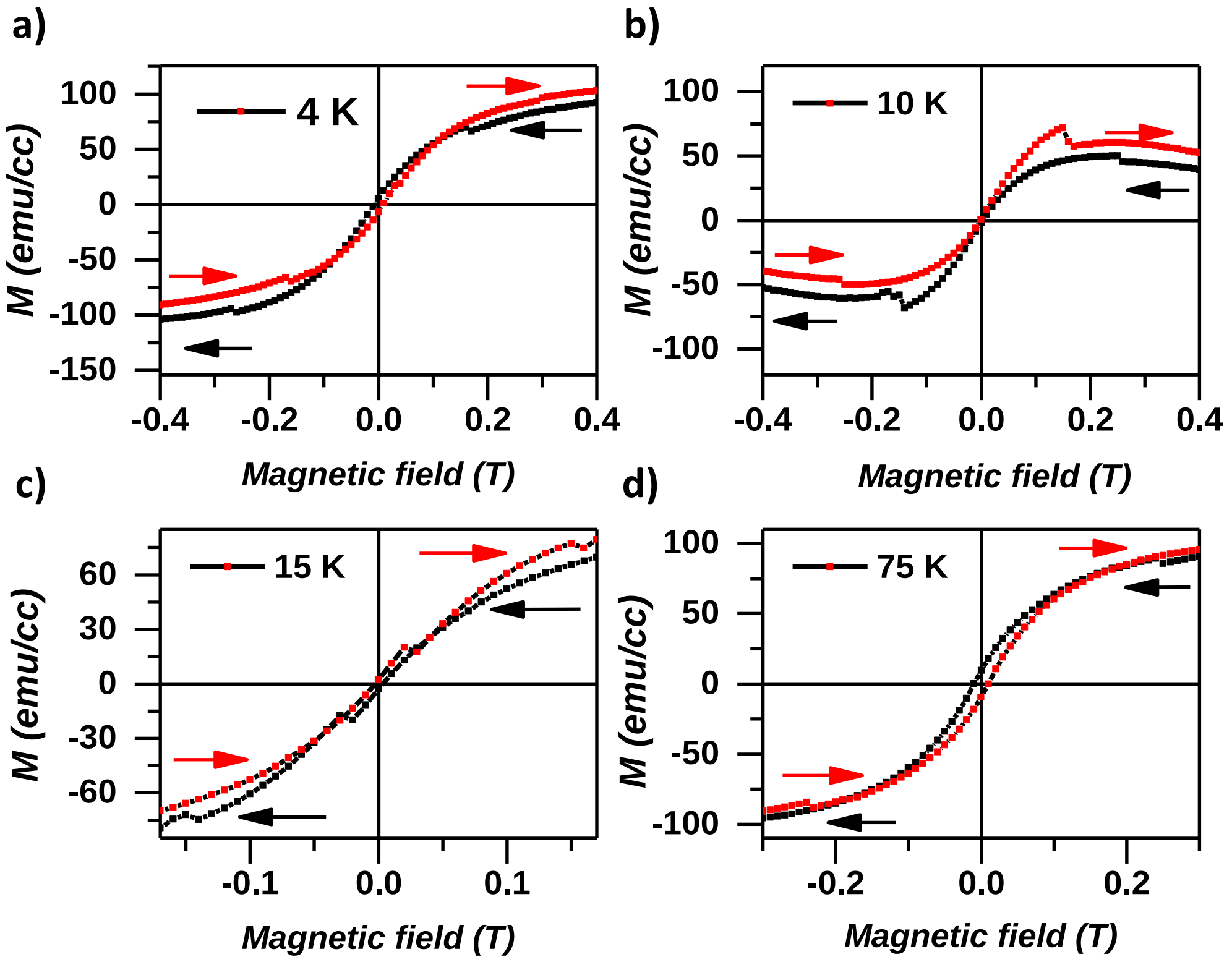

*Figure 10: Magnetization reversal of NS-SR-6 at low temperatures a) 4 K, b) 10 K, c) 15 K, and d) 75 K. The black curve and arrows represent the forward sweep of the magnetic field from +3 T to – 3 T, and the red curve represents the reverse sweep from – 3 T to 3 T.*

According to the first model, at low temperatures, the soft NSMO layer switches first, followed by the subsequent switching of the SRO layer at higher reversal fields. As the hard SRO phase retains its canted magnetization upon field reversal from saturation, the soft NSMO switches. Thus, the ferrimagnetic NSMO/SRO exchange spring shows a feeble hysteresis at 4 K. With the increase in temperature, the magnetization of individual layers varies at different rates, thus resulting in different reversal features. At 10 K, the strong AFM coupling causes the hysteresis loop to nearly close. Still, the initial switching of the soft NSMO layer leads to a small negative remanence as the field is reversed from saturation (+1 T) towards -1 T. At 15 K, the negative remanence is even more pronounced, and the magnetization loops exhibit an inverted hysteresis.

As the NSMO layer switches first at very low fields, the magnetization begins to increase in the negative direction along the applied field, leading to an inverted hysteresis and negative remanence. With a sufficient field strength in the negative direction, the SRO layers align along the field direction, causing an abrupt switching that appears as jumps in the 10 K curve. At 15 K, the switching of the SRO layer is indicated by the crossover of the reversal curves at 40 mT. The jumps represent a net reduction in magnetization upon switching the SRO layer due to the strong AFM coupling across the heterojunction, leading to an overall decrease in magnetization. This coupling is broken only at high magnetic fields, as evident from the magnetization irreversibility. Thus, the low-temperature model effectively explains the features of negative remanence and symmetric jumps in magnetization during field reversal with the ferrimagnetic exchange-spring picture.

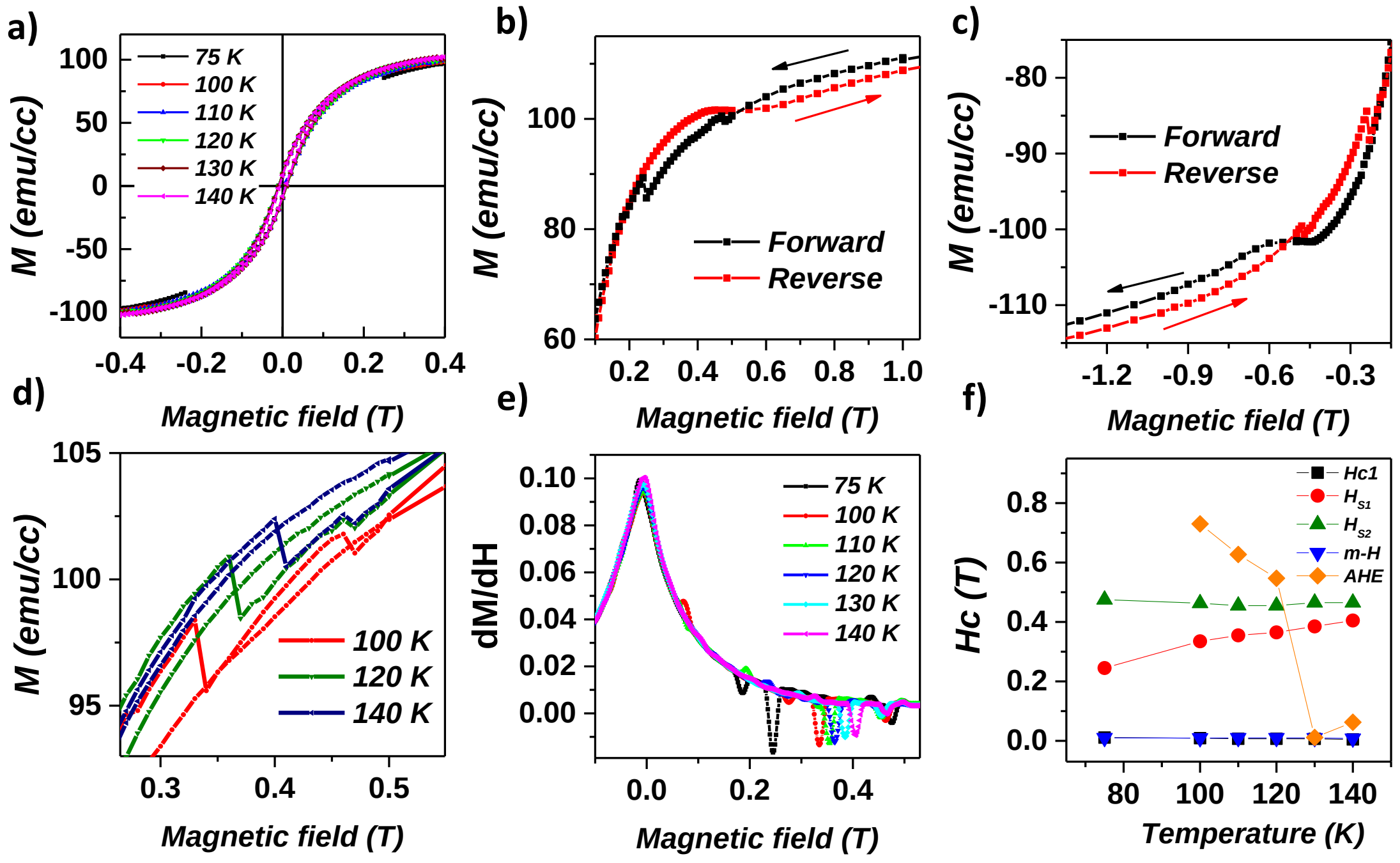

*Figure 11: a) Magnetization reversal of NS-SR-6 at temperatures > 75 K. b),c) shows the presence of multiple jumps during the forward sweep (black) from +3 T to -3 T and the plateau region during the forward sweep. Similar features are observed during the reverse sweep from -3 T to 3 T. d) Jumps in magnetization reversal at different temperatures. e) Derivative of magnetization showing multiple peaks f) Hc plot showing the coercivity and switching fields at different temperatures.*

We invoke the second model to explain the temperature dependence of magnetization at higher temperatures from 75 K to 140 K. Note that the magnetization of SRO and NSMO varies with temperature, due to which magnetization strengths of both SRO and NSMO are altered. Essentially, the SRO layer becomes softer with the increase in temperature. The schematic depicting the proposed reversal mechanism and the exchange coupling across the heterojunction is shown in Figure 12.

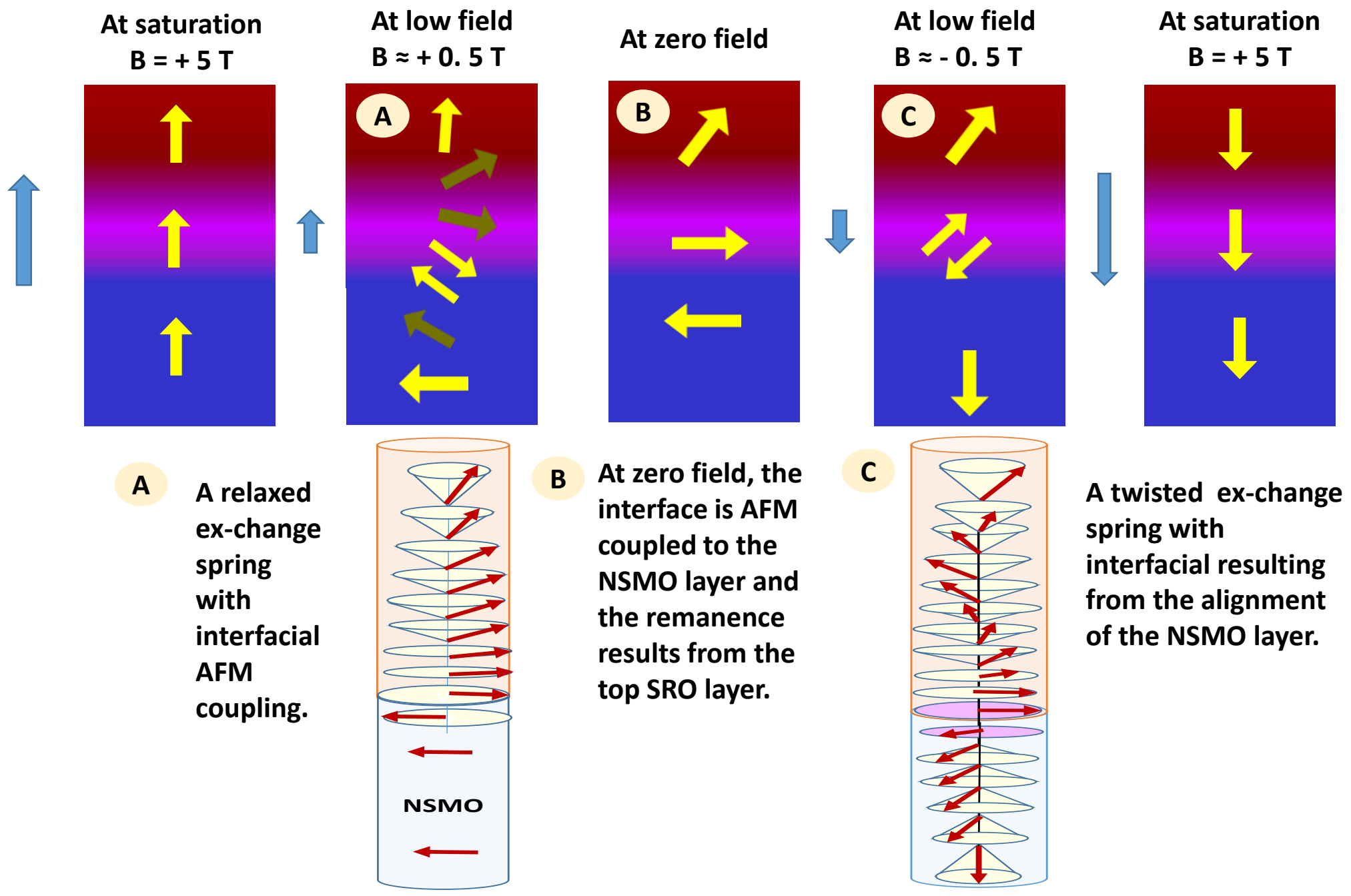

*Figure 12: Schematic representing the magnetization reversal of the NSMO/SRO heterostructure at temperatures above 75 K, where red = SRO layer at the top, Blue = NSMO layer, pink = Interfacial magnetic moments of SRO and NSMO. As the field is reduced from saturation, the magnetic moments in the system align in configurations A, B, and C, giving rise to distinct features in the magnetization hysteresis. The state A corresponds to the relaxation of the exchange spring after winding and unwinding as the field is reduced from saturation. State B corresponds to the remanent state with interfacial AFM-aligned magnetic moments. State C corresponds to the twisted exchange spring state.*

As observed from the schematic, at the saturation field, the top SRO layer, the interface, and the NSMO layer align along the field direction due to large Zeeman energy. Upon reduction of the field to ~ 0.5 T, the top SRO layers tend to retain their canted magnetization along its MCA (the out-of-plane direction). As the field is reduced further, the NSMO layer flips. Subsequently, the interfacial SRO aligns anti-ferromagnetically with NSMO, resulting in the unwinding of the SRO moment and thereby causing an increase in magnetization. The Ru-moments at the interface have a component/projection along the relaxed Ru-moments in the top layer. Thus leading to an overall structure with a net increase in magnetization, emerging as jumps at +0.48 T and +0.25 T. Multiple jumps are present in the positive quadrant, pointing to the reorientation of exchange coupled springs. At zero field, the observed hysteresis arises due to net magnetization of the exchange spring, with a significant contribution from the relaxed top layer SRO while interfacial SRO and NSMO moments are coupled in AFM. Upon further increase in the field towards negative fields, the NSMO layer tends to align along the perpendicular field direction, resulting in an interesting exchange spring configuration. Let's call this field the switching field $H_s$. For fields 0 T > H > $H_s$, the top SRO layers with strong MCA remain in the remanent state. As the NSMO layer aligns with the field, a twist is established in the exchange spring because the interfacial Ru moments must maintain AFM coupling. Thus, a twist in magnetic moments is established at $H_s$, as depicted in schematic Figure 12 c). This twist propagates through the thickness of the SRO layer, resulting in a twisted, non-collinear exchange spring. In addition, the signature of such a twist is observed as a plateau region in magnetization reversal from -0.4 T to -0.6 T and as a crossing of the magnetization reversal curves at -0.5 T at 75 K. Such a twisted exchange spring has been previously reported by S. Das et al. in the LSMO/SRO heterostructure [17]. They interpreted the formation of an exchange spring network resembling a Bloch wall in SRO induced by the AFM coupling at the interface[17]. With further reversal of the field beyond -0.5 T, the Zeeman energy in the heterostructure increases, leading to the saturation of the magnetization. This results in unwinding the exchange spring and the eventual breaking of AFM coupling at high fields. During field reversal from negative saturation, similar features of jumps and plateaus are observed.

Subsequently, with increased temperature, the fields at which the jumps emerge ($H_{peak}$) shift towards higher fields. As the field is reduced from saturation, the exchange spring rearrangement occurs at smaller ΔH ($\Delta H = H_{saturation} - H_{peak}$) with increased temperature. This shows that the stiffness of the exchange spring coupling reduces as temperature increases due to a reduction in sub-network magnetization, effectively giving rise to jumps at smaller ΔH. The plateau region becomes less distinct as the measuring temperature increases to 140 K. Thus, the delicate balance between the magnetocrystalline anisotropy, interfacial exchange coupling energy, and Zeeman energy governs the winding, unwinding, and the strength of the exchange spring. Therefore, a non-collinear twisted exchange spring with progressive canting along the out-of-plane direction manifests at the switching field at temperatures around 75 K to 140 K.

The magnetization behavior of the NS-SR-23 with intermediate thickness of SRO is thoroughly investigated based on the proposed magnetization reversal models. We briefly discuss the magnetization behavior of the thinnest and thickest heterostructure, NS-SR-6 and NS-SR-80. The NS-SR-6 sample shows weak hysteresis ferromagnetic hysteresis loops over the range of 4 K to 75 K. At 4 K, the hysteresis has a remanence of ~ 27 emu/cc with a coercivity of 19 mT in NS-SR-6. We consider the reversal curve from + 3 T, where the field is reversed continuously along the negative direction. The presence of jumps in magnetization is noted close to -1 T, resulting in a net reduction in magnetization. It is observed that no saturation is reached even at 3 T field. Upon reversal of the magnetization from – 3 T, the magnetization irreversibility is present however, the sweeps do not have any cross-overs. A symmetric drop in magnetization is observed at +1 T upon reversal from -3 T.

We can understand the reversal behavior based on our low-temperature model. The NSMO layer having higher magnetization gives rise to the total FM character. Though the NSMO layer is soft, unlike in NS-SR-23, the NSMO layer doesn't switch at positive fields during reversal. This is because of the influence of the bulk-like SRO layer at the top, which retains its net magnetization along its MCA axis, which is very feeble due to its low thickness compared to NS-SR-23. Therefore, at zero field, the NSMO layer and the uncoupled SRO layers relax along their MCA, leading to a remanence despite the AFM coupled interface. Upon further reversal, the NSMO layer switches near its coercive fields. However, to maintain the AFM coupling, the SRO layers align AFM, resulting in switching and a net magnetization drop. This is evident as a small jump at -1 T. The AFM interfacial coupling is again apparent from the hysteresis openings/magnetization irreversibility while changing the sweep direction. Note that the switching field is significantly high here (~ 1 T). This is because of the increased strength of AFM coupling in the thinnest bilayer. In contrast, in the thickest heterostructure (NS-SR-80), we observe (i) a two-component magnetization without additional jumps and (ii) high remanence of ~ 200 emu/cc and high coercivity -0.63 T at 4 K. The two-component magnetization is clearly discernable at higher temperatures (75 K). The larger coercivity and the two-component magnetization show that the extent of exchange spring coupling (Bloch wall width) increases in the thicker heterostructure.

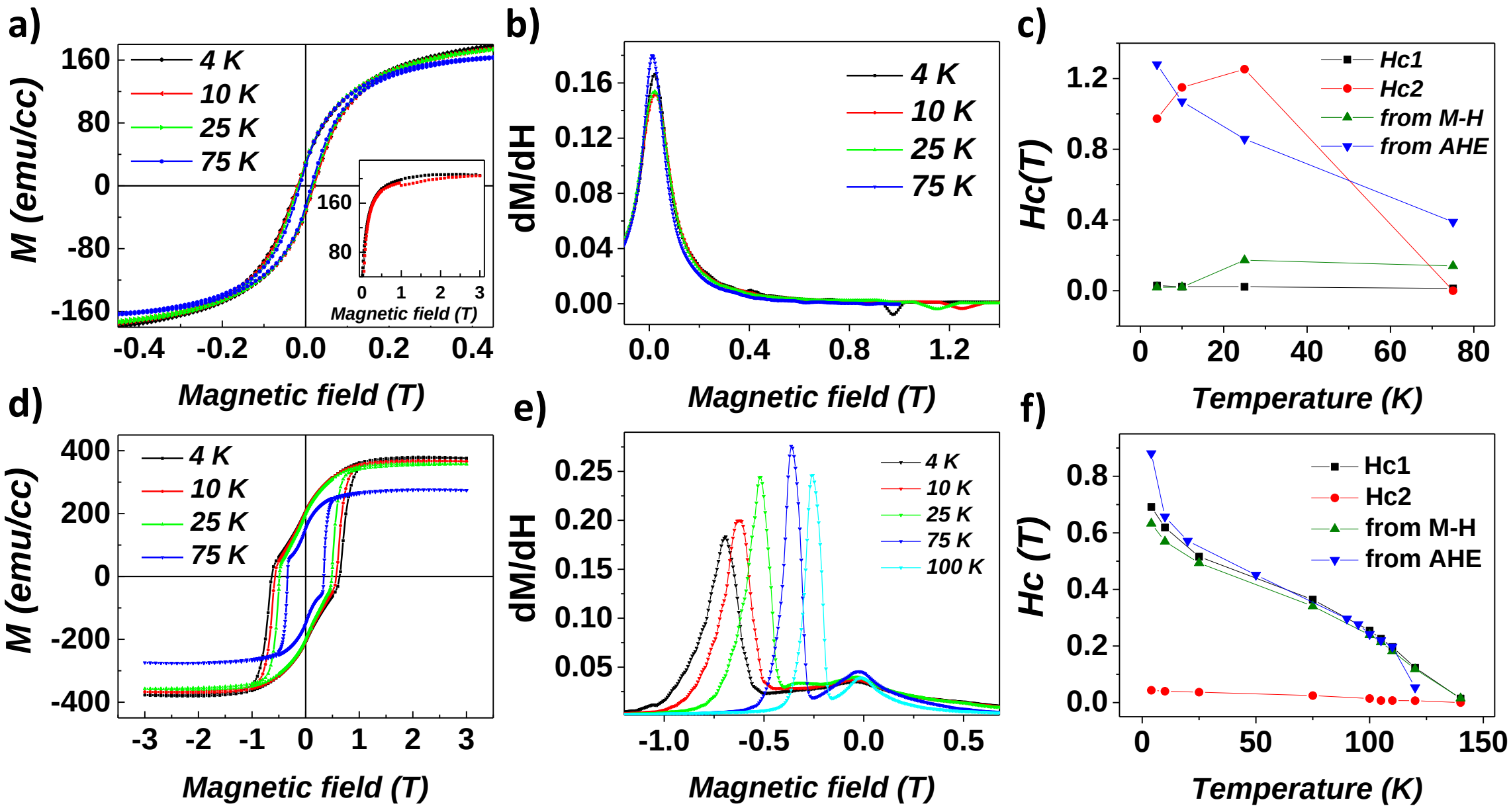

*Figure 13: Magnetization behavior of thinnest and thickest NSMO/SRO heterostructure a)-c) NS-SR-6 and d)-f) NS-SR-80. a) Magnetization hysteresis of NS-SR-6 showing magnetization irreversibility during field reversal at low-temperature as depicted in inset b) Magnetization derivative plot with central peak and humps at higher fields arising from switching of magnetization sub-network c) Coercivity plot d) Magnetization hysteresis of NS-SR-80 e) Magnetization derivative plot f) Hc analysis.*

## 4. Discussion:

Previous studies have shown the emergence of additional dips/humps in AHE near the switching/coercive fields and attributed them to the presence of non-trivial spin texture present in the system[35]. Several studies have also clarified the THE arising in ultra-thin SRO films due to defects such as variation in thickness in the order of unit cells [9,36,37].

Such features are often argued with the famous two-channel anomalous hall transport, where magnetically separate regions possess different AHE coefficients that change sign over a wide temperature range [38]. This is the case with Roy et al.'s work, where two thin films of SRO with thicknesses 14 and 24 nm are studied, with one showing THE-like features [9]. However, the two-channel AHE was ruled out as the field of second AHEs could not be correlated with $H_c$. Therefore, the presence of THE has been attributed to structural distortions in ultra-thin 14 nm SRO film, which shows THE [38]. Later, the THE feature is explained by employing diffusive berry phase transition and exchange-coupled magnetic phases with different coercivity and Berry phases [9]. The hard magnetic phase has a sharp Berry phase transition, while the soft one has a diffusive Berry phase transition[9]. In our case of a single layer 23 nm thick SRO thin film, we see the presence of two magnetic phases from the derivative plot. It is also observed that the one-order difference in the coercivity of the two phases indicates a hard and soft magnetic phase of SRO within the thin film, which is why a two-step magnetization reversal is evident. The soft magnetic phase in SRO has been attributed to the tetragonal SRO system with reduced MCA, whereas the orthorhombic SRO system has enhanced MCA properties [39]. The tetragonal phase may result from lattice volume expansion arising from the oxygen and Ru deficiencies and strain, leading to suppression of magnetic anisotropy at the interface [15]. The SRO layer closer to the substrate forms a highly strained interface however, away from the interface, the strain can relax. Thus, the hard and soft magnetic phases may arise in the SRO single layer due to strain, leading to a two-step magnetization reversal. We presume that though there are hard and soft phases of SRO, the coupling among the phases is relatively weak, due to which there may not be any magnetic non-collinearity. Thus, we do not see THE-like signatures in the AHE.

However, the presence of THE-like features in NSMO/SRO heterostructures clearly indicates the effect of magnetic proximity arising from the NSMO layer. From our magnetization models, we have established that a non-collinear exchange spring is present within the system. However, the occurrence of THE-like feature only close to the AHE-crossover temperature remains a question. In previous studies on other manganites/ruthenate heterostructure, LCMO/SRO, Roy et al. attributed the emergence of THE-like feature to the presence of two magnetic phases in heterostructure: hard and soft with different $H_c$s where one of the $H_c$ corresponding to the hard phase-matched closely to the peaks in MR (coercivity in M-H) [40]. The $H_c$s are mapped to the two berry phases present within the system, and the diffusive berry phase transition is used to explain the occurrence of THE-like feature in LCMO/SRO[40]. We also carried out similar $H_c$ analyses and found that multiple $H_c$s exist in the system. The $H_c$ analysis is performed by taking the derivative of the magnetization loops as shown in Figure 13 b), e) (we consider a single sweep from positive to negative saturation), and the presence of multiple switching fields is confirmed. The $H_c$ plots are shown in the Figure 13 c), f). Comparing the values of $H_c$ deduced from magnetic (M-H) and transport measurements (AHE), their temperature dependence shows no correlation with the field at which THE-like feature is observed in AHE.

Therefore, the establishment of exchange-spring-like coupling in the NSMO/SRO system and their reversal mechanism close to higher temperatures (75 K to 140 K) is believed to play a decisive role in the anomaly observed in the AHE loops. A closer inspection of the AHE feature in the NS-SR-23 reveals that the THE-like hump at 120 K is obtained at -0.2 T only upon the reversal of the magnetic field from saturation. This closely correlates to the establishment of twist after switching the NSMO layer, as depicted in the reversal mechanism, Figure 12 c). The plateau region and the cross-over of reversal curves at 75 K are evident in the field range -0.2 T to – 0.5 T, and this closely matches with the field of -0.2 T at which THE feature occurs. As explained in the reversal model, the presence of non-collinear twist is decisive for THE like feature. However, the bulk-like SRO layer, far from the interface, can remain hard due to high MCA. The reduction in magnetization with temperature may aid in establishing twisted exchange coupled moments across the SRO's thickness. The magnetotransport, being dominated only by the SRO layer, shows only AHE at temperatures less than Tc of SRO. However, with an increase in temperature, the establishment of the non-collinear spin structures in NSMO/SRO leads to the emergence of THE-like features in AHE close to Tc.

## 5. Conclusion:

The magnetic and magnetotransport properties of NSMO/SRO heterostructures are studied in this work. The growth parameters were effectively tuned to obtain uniform and homogeneous SRO thin films on $TiO_2$-terminated STO substrates with varying thicknesses. A two-step magnetization reversal in SRO thin films indicates the presence of a soft interfacial strained SRO layer and an unstrained hard magnetic phase. No additional dips or peaks were observed in the anomalous hall transport of ultrathin SRO films of 6 nm. However, the anomalous Hall measurements on NSMO/SRO heterostructures show hump-like features originating from the proximity effect of NSMO. The NSMO/SRO heterostructures with varying SRO layer thicknesses show systematic relaxation in compressive strain with increasing SRO layer thickness. Magnetotransport investigations revealed the presence of humps and dips around the temperature at which the AHE changes sign. The emergent humps in AHE persist even in thicker heterostructures. From magnetization studies, it was understood that an AFM exchange coupling exists between the interfacial SRO and NSMO, forming an exchange spring. The subtle interplay between the temperature dependence of magnetization, MCA, and Zeeman energy drives the reversal mechanism in the exchange spring in two configurations: relaxed and twisted states. The emergent twisted exchange spring results in non-collinearity of magnetic moments across the heterojunction thus giving rise to THE-like features in NSMO/SRO heterostructures.

## Acknowledgements:

We would like to acknowledge the Department of Atomic Energy, India for the provision of experimental facilities. We thank UGC-DAE CSR, Kalpakkam node, for providing access to magnetic and magnetotransport measurement systems. The authors are grateful to RRCAT, Indore, for beam line facilities.